\documentclass[showpacs,pre,twocolumn,amsmath,amssymb,epsfig]{revtex4-1}

\usepackage{graphicx}
\usepackage{epsfig}
\usepackage{dcolumn}% Align table columns on decimal point
\usepackage{bm}
\newcommand     {\beq}[1]         { \begin{equation} #1 \end{equation} }
\newcommand     {\beqa}[1]        { \begin{eqnarray} #1 \end{eqnarray} }

\begin{document}

\title{Record breaking bursts during the compressive failure of porous 
materials} 

  \author{Gerg\H o P\'al${}^{1}$, Frank Raischel${}^{1}$, Sabine Lennartz-Sassinek${}^{2,3}$,
  Ferenc Kun${}^{1}$\email{ferenc.kun@science.unideb.hu}, 
  Ian G.\ Main${}^{2}$} 
 \affiliation{$^1$Department of Theoretical Physics, University of Debrecen,
 P.O. Box 5, H-4010 Debrecen, Hungary}
 \affiliation{$^2$School of Geosciences, University of Edinburgh,  EH9 3JL
 Edinburgh, UK}
 \affiliation{$^3$Institute for Geophysics and Meteorology, 
 University of Cologne, Cologne, Germany}

\begin{abstract}
An accurate understanding of the interplay between random and
deterministic processes in generating extreme events is of critical
importance in many fields, from forecasting extreme meteorological
events to the catastrophic failure of materials and in the Earth.  Here
we investigate the statistics of record-breaking events in the time
series of crackling noise generated by local rupture events during the
compressive failure of porous materials. The events are generated by
computer simulations of the uni-axial compression of cylindrical samples
in a discrete element model of sedimentary rocks that closely resemble 
those of real experiments. The number of records
grows initially as a decelerating power law of the number of events,
followed by an acceleration immediately prior to failure. The
distribution of the size and lifetime of records are power-laws with
relatively low exponents. We demonstrate the existence of a
characteristic record rank $k^*$ which separates the two regimes of the
time evolution. Up to this rank deceleration occurs due to the effect of
random disorder. Record breaking then accelerates towards macroscopic
failure, when physical interactions leading to spatial and temporal
correlations dominate the location and timing of local ruptures. 
Scaling analysis revealed that the size distribution of records of different 
ranks has a universal form independent of the record rank. 
Sub-sequences of bursts between consecutive records
are characterized by a power law size distribution with an exponent which 
decreases as failure is approached. High rank records are 
preceded by bursts of increasing size and waiting time between consecutive events
and they are followed by a relaxation process. 
As a reference, surrogate time series 
are generated by reshuffling the crackling bursts. The record statistics of 
the uncorrelated surrogates agrees very well with the corresponding 
predictions of independent identically distributed random variables, which confirms
that the temporal and spatial correlation of cracking bursts are responsible for the 
observed unique behaviour.
In principle the results could be used to improve forecasting of
catastrophic failure events, if they can be observed reliably in real
time.

\end{abstract}

\maketitle

\section{Introduction}
The compressive failure of heterogeneous materials proceeds in bursts 
of cracking events. Measuring the generated acoustic emissions is 
the primary source of information about the time evolution of the 
fracture process \cite{sammonds_role_1992,ojala_2004,Mair200025,Henderson1992905,
rosti_crackling_2009,nataf_avalanches_2014,vives_0953-8984-25-29-292202}. 
The statistical analysis of the stochastic 
time series of crackling bursts in field data, laboratory experiments, 
and in computer simulations has provided a useful insight into the 
accumulation of damage and into the approach of the system 
to macroscopic failure. The ultimate challenge of the field is to find
statistical signatures which could be exploited to forecast the impending 
catastrophic failure \cite{ROG:ROG1468}. For this purpose the analysis of synthetic time 
series of simulated fracture processes is indispensible since they 
allow a range of variables to be controlled and investigated independently, 
and allow representative sampling of underlying trends and statistical 
variability over a large number of trials  
\cite{alava_statistical_2006,hidalgo_avalanche_2009}. 

Recently, we have introduced a discrete element model of porous sedimentary 
rocks which captures the essential ingredients of the materials' 
micro-structure and of the dynamics of breaking 
\cite{PhysRevE.88.062207,PhysRevLett.112.065501}. 
The model was used to investigate the compressive failure 
of cylindrical samples under strain controlled conditions. During the failure 
process we identify cracking bursts as correlated trails 
of breaking beams which are 
generated due to the gradual stress redistribution in the sample following 
local failure events \cite{PhysRevE.88.062207,PhysRevLett.112.065501}. 

In the present paper we investigate the internal structure of the time series 
of breaking bursts by analyzing the statistical features of record breaking 
(RB) 
events. Records are bursts which have a size greater than any previous
events of the series. Recently, the record statistics of stochastic time series
has attracted great attention due to its relevance for climate and earthquake
research 
\cite{wergen_krug_0295-5075-92-3-30008,GRL:GRL21495,PhysRevE.77.066104,npg-17-169-2010}. 
Interesting analytic results have been obtained for the RB statistics
of the sequences of independent identically distributed (IID) random variables
for a randomly-sampled stationary process
\cite{wergen_record_2013,PhysRevE.87.052811,npg-17-169-2010}. 
In physics the statistics of records 
has been applied to understand correlated processes emerging in various types 
of random walks 
\cite{PhysRevLett.101.050601,1751-8121-45-35-355002,1751-8121-47-25-255001},
superconductors \cite{PhysRevB.71.104526}, domain wall dynamics in spin glasses 
\cite{records_complex_system}, 
and in chaotic processes \cite{srivastava_records_2015}. 
The record statistics 
of the bursting activity has also been studied recently in models exhibiting 
self organized 
criticality (SOC) \cite{PhysRevE.87.052811} and in a mean field model of 
fracture \cite{danku_frontiers_2014}. For earthquakes record statistics were recently
applied to reveal spatiotemporal clustering of seismicity either by focusing 
on the inter event times \cite{npg-17-169-2010} or using both the spatial 
and temporal distance of events \cite{GRL:GRL21495,PhysRevE.77.066104}.
In our realistic discrete element model
of compressive failure we analyze both the aggregated statistics 
of records and the evolution of record breaking as the sample approaches 
macroscopic failure. 
As a null hypothesis we compare our results to the IID findings 
and to the record statistics of a surrogate data set where correlations 
are destroyed by randomly reshuffling the breaking bursts of fracture simulations.
This comparison makes it possible to reveal interesting trends and correlations 
in the spatial and temporal signature of the crackling noise.
We show significant departures associated with non-stationary processes 
associated with increased strain, and reveal new signatures of impending 
catastrophic failure in the time series associated with record-breaking events. 
 
\section{Record breaking events}
We study the statistics of records in a synthetic time series of breaking bursts
generated by a discrete element model (DEM).
The model has been developed recently to investigate the emergence of crackling 
noise
during the compressive failure of cylindrical samples of porous rocks 
\cite{PhysRevE.88.062207,PhysRevLett.112.065501}.
 The rock sample was reconstructed
on the computer by sedimenting spherical particles with a realistic size 
distribution.
The particles are coupled by cohesive contacts represented by beam elements 
which break when they are stressed beyond a limit. 
Strain controlled uniaxial compression was realized by clamping a few particle 
layers 
at the bottom and at the top of the sample. The bottom was fixed, while the top 
layers
were moved at a constant speed along the cylinder axis. The loading process was 
stopped 
when the force acting on the top layer dropped down to zero.
Due to the subsequent load redistribution
following failure events beams break in cascades analogous to crackling 
avalanches in real 
materials. In Refs.\ \cite{PhysRevE.88.062207,PhysRevLett.112.065501} we 
investigated the 
dynamics of emergence and statistics of such crackling bursts during the strain 
controlled
uniaxial compression of cylindrical samples composed of 20000 particles. 
The modeling approach proved to be successful in reproducing several important 
observed features of crackling noise in porous materials 
\cite{PhysRevLett.110.088702,nataf_avalanches_2014,vives_0953-8984-25-29-292202,
sammonds_role_1992,ojala_2004,Mair200025}.

\begin{figure}%[!h]
 \begin{center}
 \epsfig{ bbllx=20,bblly=10,bburx=410,bbury=440,file=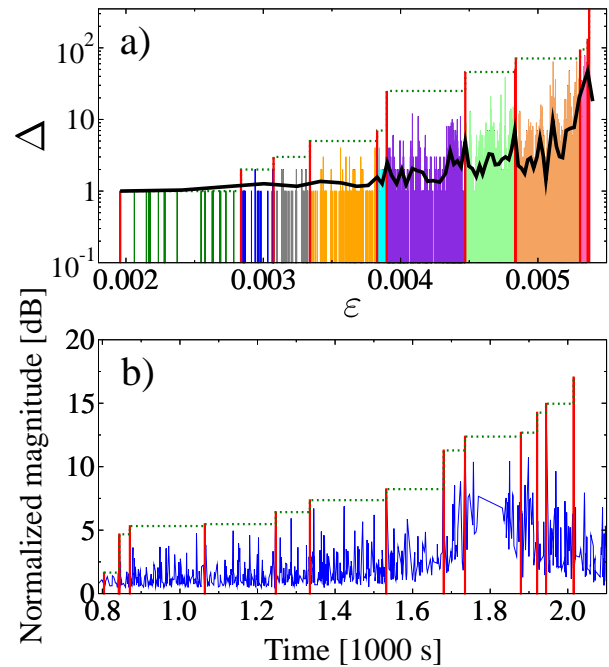,
 width=8.3cm}
 \end{center}
 \caption{%\small 
 (Color online) $(a)$ Sequence of breaking bursts in a single simulation, i.e.\ 
 the size of events $\Delta$ as a function of the strain $\varepsilon$ where 
they occurred.
 Record events are highlighted by red bars, while sub-sequences between records
 have randomly assigned colors.
 The bold continuous line represents the moving average of burst sizes using
 30 consecutive events. For clarity, a relatively small system of 20000 particles
 was considered, where $n_{max}=1832$ bursts occurred with 10 records.
 $(b)$ Record events in a time series of a real fracture experiment performed 
 under conditions similar to the simulations \cite{sabine_ian_pre_2014}.
 }
 \label{fig:series}
 \end{figure}

Figure \ref{fig:series} demonstrates the breaking sequence of a single simulation
of a system of 20000 particles where 1832 bursts are obtained 
up to macroscopic failure in comparison with data from a real experiment \cite{sabine_ian_pre_2014}. 
In the example the breaking process starts 
at the deformation $\varepsilon\approx 0.0019$ 
where initially small bursts occur with size $\Delta=1$. As loading proceeds 
larger and larger bursts are triggered so that the average burst 
size steadily increases towards failure although $\Delta$
has strong fluctuations due to the disordered micro-structure of the
material. 
 
Records of the event series are bursts which have a size $\Delta$ greater 
than any previous event. RB events are identified by their rank $k=1,2, \ldots$ 
which occurred as the $n_k$th event of the complete time series with size 
$\Delta_r^k$.
By definition the first event $n=1$ of the series is a record so that $n_1=1$ 
holds.  
Figure \ref{fig:series} illustrates that RB bursts form a monotonically 
increasing
sub-sequence and divide the time series into segments of varying number 
of smaller size events. The presence of RB events
has a complex effect on the local structure of the time series in Fig.\ \ref{fig:series}:
the moving average of the event size tends to peak at the time of 
record-breaking events, with a precursory increase in the average burst size, 
followed by a decrease or relaxation after the record-breaking event.  This pattern is
more pronounced close to failure.

To characterize how record breaking evolves during the loading process,
we also consider the size increments $\delta_r^k$ and the waiting times $m_k$ 
between consecutive records defined as
\beqa{
\delta_r^k &=& \Delta_r^{k+1} - \Delta_r^k, \label{eq:increm} \\
m_k        &=& n_{k+1} - n_k, \label{eq:mk} 
}
respectively. 
After analyzing the overall statistics of record quantities, 
we focus on the evolution of the sequence of RB events 
as the system approaches failure. Finally, we study how records influence
the structure of the time series of crackling events.

 \begin{figure}%[!h]
 \begin{center}
 \epsfig{ bbllx=40,bblly=30,bburx=370,bbury=330,file=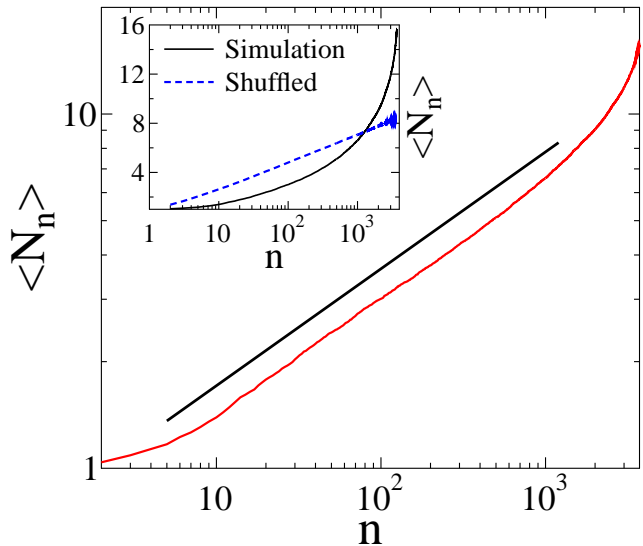,
 width=8.5cm}
 \end{center}
 \caption{%\small 
 (Color online) Average number of records $\left<N_n \right>$ occurred until the 
$n$th burst as a function of $n$ on a double logarithmic plot. Power law behavior is 
obtained over a broad range of $n$. The straight line represents a power law of exponent 
$0.33$. The inset presents the same quantity $\left<N_n \right>$ on a semi-logarithmic plot.
For comparison the shuffled data set is also included where a logarithmic dependence
is found.
 }
 \label{fig:recnum}
 \end{figure}
Based on the statistics of extremes it has been shown for sequences of 
independent 
identically distributed random variables (IID) that the statistics of 
record breakings has universal features, i.e.\ several statistical measures of 
IID records do not depend on the underlying probability distribution of individual 
events \cite{wergen_record_2013}.
The increasing average event size and decreasing waiting time between 
consecutive 
events in Fig.\ \ref{fig:series} demonstrate that the time series of crackling 
bursts accompanying compressive failure is 
highly non-stationary. Comparing the results of our analysis to the 
corresponding 
results of IIDs enables us to quantify the competing role 
of the structural disorder of the material and of the stress enhancements around 
failed
regions, which favorably lead to a stationary sequence of uncorrelated events 
\cite{main_damage_2000,Mair200025,Heap201171,main_fault_2000} and to an
accelerating sequence with spatial and temporal correlations 
\cite{PhysRevLett.112.065501,
PhysRevE.88.062207,Mair200025,Heap201171,main_fault_2000}, respectively. 
Additionally, we generate a surrogate data set by reshuffling the events of the 
simulated time series to destroy correlations. For each fracture simulation the 
indices $n$ of events are permutated and then the same analysis was performed 
as for the original data. When it is applicable, the results of the original data,
its suffled counterpart, and the IID predictions are presented together in the figures.
For the data evaluations the statistical toolbox of MATLAB was extensively used
\cite{MATLAB:2010} where a non-linear least square method was used for curve fitting.

In the present study careful parallelization of the computer code allowed us to 
substantially
increase the system size. Here simulations were performed with a particle 
number fluctuating around $10^5$ such that on the average the diameter 
of the cylinder is spanned by about 50 particles while the height to 
diameter aspect ratio of the sample is 2.3 \cite{PhysRevE.88.062207,PhysRevLett.112.065501}.
For an average particle size of 200 microns (typical reservoir sandstone) 
the sample diameter would correspond to 1cm which gets near to the typical 
small core lab sample diameter of 2.5 cm. In a single simulation we identified 
$n_{max}=3500-4000$ bursts of beam breakings. Averages of all quantities are 
calculated over 550 simulations.

\section{Number of records}
Since records are major bursts of the breaking process which have a 
dominating contribution to the accumulating damage, 
it is of high interest how the number of records
$N_n$ increases with the number of events $n$. 
For IIDs it has been shown that the average number of records 
$\left<N_n\right>$ 
that occurred until the event number $n$  has been reached grows 
logarithmically with $n$ 
\beq{
\left<N_n\right> \sim \kappa + \ln n + \mathcal{O}(1/n), \qquad \mbox{for} \qquad n\to +\infty,
\label{eq:numrec_log}
}
independently of the specific form of the probability density of the random 
variables \cite{wergen_record_2013}. Here $\kappa$ denotes the Euler-Mascheroni constant
$\kappa\approx 0.577215665$ \cite{wergen_record_2013,PhysRevE.87.052811}.
For crackling noise accompanying the quasi-static loading of heterogeneous 
materials deviations from this behavior can be expected: due to the increase 
of the externally imposed strain the failure process accelerates so that 
larger bursts are triggered which follow each other after shorter waiting times 
\cite{PhysRevE.88.062207,PhysRevLett.112.065501}. The complex redistribution
of stress inside the damaged sample gives rise to the emergence of temporal and 
spatial correlations of bursts of the sequence. As a consequence, 
the number of records increases as a power law of the event number
\beq{
\left<N_n\right> \sim n^{\alpha},
\label{eq:averrwcnum}
}
over a broad range of $n$ (see Figure \ref{fig:recnum}). 
The value of the exponent was obtained numerically $\alpha=0.33\pm 0.03$. 
This value of $\alpha$ indicates that the number of record-breaking events also increases 
at a decelerating rate, albeit with a different form to that of Eq.\ (\ref{eq:numrec_log}). 
Deviations from the power law can be observed at the 
very beginning of the loading process and in the close vicinity of macroscopic 
failure.
The up-turn of the $\left<N_n\right>$ curve in Fig.\ \ref{fig:recnum} 
for the highest event indices $n$ shows that as 
failure is approached the rate of record-breaking accelerates 
(the local slope increases). This acceleration generally occurs 
when the record number $N_n$ exceeds 7. The inset of Fig.\ \ref{fig:recnum}
presents $\left<N_n\right>$  on a semi-logarithmic plot. Strong deviation 
can be observed from a straight line which confirms that the functional form
is not logarithmic.
 \begin{figure}%[!h]
 \begin{center}
 \epsfig{ bbllx=30,bblly=20,bburx=370,bbury=330,file=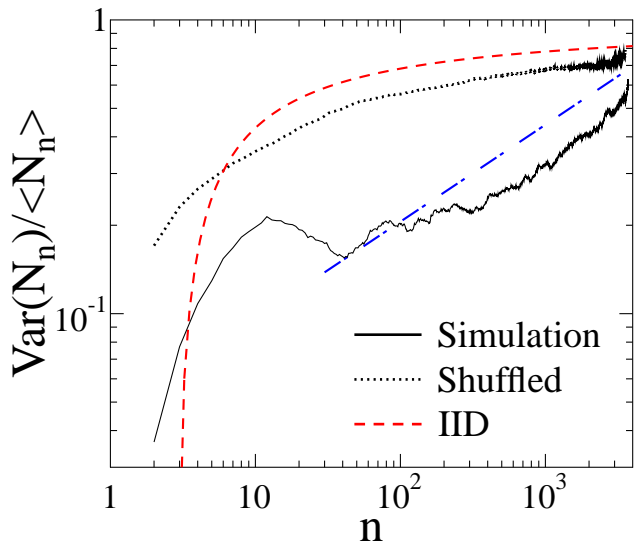,
 width=8.5cm}
 \end{center}
 \caption{%\small 
 (Color online) Standard deviation of the number of records $Var(N_n)$ occurred until
 the $n$th burst divided by the average record number $\left<N_n\right>$ 
 on a double logarithmic plot. The red (dashed) and the blue (dashed-dotted) lines
 represent the corresponding result of the shuffled data set and a power law of 
 exponent 1/3, respectively. 
 }
 \label{fig:recnum_std}
 \end{figure}
The corresponding curve of the shuffled data is also presented in the inset, which
has a perfect agreement with the analytical IID result Eq.\ (\ref{eq:numrec_log}).
From Figure \ref{fig:recnum}$(inset)$ the null hypothesis of sampling from a parent distribution 
can be rejected. Instead we have a decelerating transient response of a power-law form.

To characterize the sample-to-sample variation of the number of records $N_n$ we 
calculated the standard deviation 
\beq{
Var(N_n)=\left<(N_n-\left<N_n\right>)^2\right>.
}
For IIDs the analytic solution gives again an asymptotic logarithmic increase with $n$ 
as \cite{wergen_record_2013,PhysRevE.87.052811}
\beq{
Var(N_n)=\kappa + \ln n -\frac{\pi^2}{6} + \mathcal{O}(1/n), \quad \mbox{for} \quad n\to +\infty,
}
so that the relative variance $Var(N_n)/\left<N_n\right>$ monotonically increases and
tends to 1 for sufficiently large $n$. Figure \ref{fig:recnum_std} presents that 
the very beginning of our fracture process, up to about $n=10$ bursts, 
is consistent with the IID behaviour. However, after a short intermediate decreasing
regime the curve sets to a faster increase which has a nearly power law functional form.
In Fig.\ \ref{fig:recnum_std} a power law of the same exponent as for the average record
number in Fig.\ \ref{fig:recnum} is drawn to guide the eye. The result of the shuffled data
is again consistent with the IID solution.
 
\begin{figure}%[!h]
 \begin{center}
 \epsfig{ bbllx=30,bblly=20,bburx=370,bbury=330,file=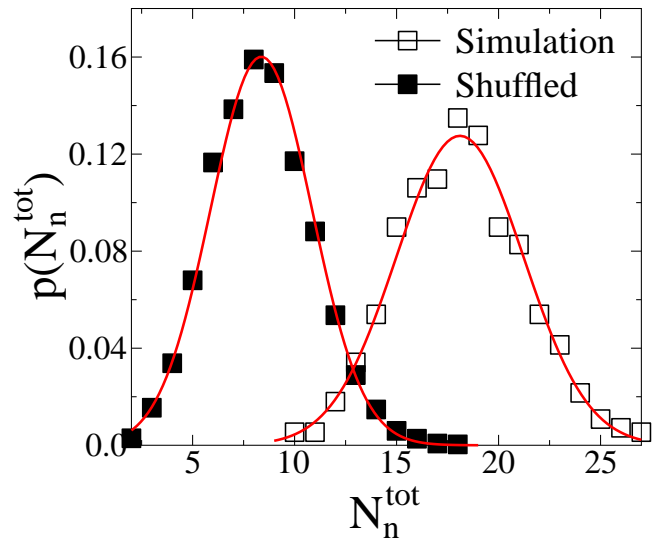,
 width=8.3cm}
 \end{center}
 \caption{%\small  
 (Color online)
 Probability distribution of the total number of records $p(N_n^{tot})$ 
 that occurred up to failure for the original and surrogate data sets.
 The continuous lines represent the Gaussian distributions obtained with the mean and standard 
 deviation calculated directly from the data.
 }
 \label{fig:recnumdist}
 \end{figure}

The probability distribution of the number of records $p(N_n)$ that occurred 
up to a fixed number $n$ of events has been found to have a Gaussian functional
form for IIDs in the limit of large $n$ values 
\cite{wergen_record_2013,PhysRevE.87.052811}. 
Figure \ref{fig:recnumdist} shows that in our fracture process
the distribution $p(N_n^{tot})$ of the total number of records $N_n^{tot}$ that 
occurred 
up to failure is similar to a Gaussian with an average $\left<N_n^{tot}\right>=18.1$ 
and standard deviation $Var(N_n^{tot})=3.1$. 
The highest and lowest number of records 
we identified in single simulations were 10 and 26, respectively, with the most 
probable value 18. 
In the shuffled event series large size events can occur anywhere which results
in a significantly lower number of records $\left<N_n^{tot}\right>=8.4$ with a 
standard deviation $Var(N_n^{tot})=2.5$. The continuous lines in Fig.\ 
\ref{fig:recnumdist} represent Gaussian distributions obtained by inserting the corresponding 
values of $\left<N_n^{tot}\right>$ and $Var(N_n^{tot})$ for the two data sets.
Both for the simulated and surrogate data reasonable agreement is obtained 
with the Gaussian distribution.

\section{Distribution of record sizes}
\begin{figure}%[!h]
 \begin{center}
 \epsfig{ bbllx=50,bblly=20,bburx=380,bbury=330,file=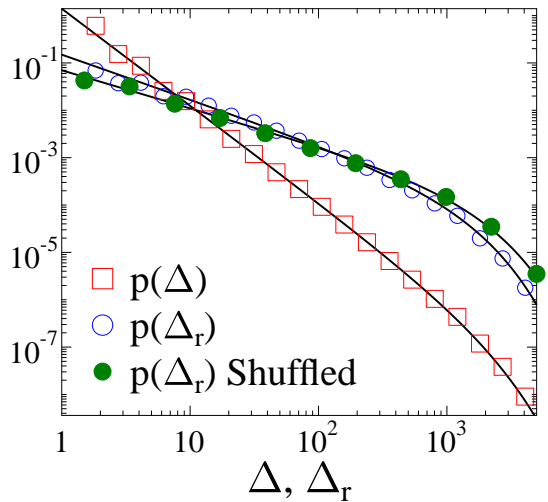,
 width=7.5cm}
 \end{center}
 \caption{%\small 
 (Color online) Probability distribution $p(\Delta)$ of the burst size $\Delta$ 
 considering all events up to failure. 
 The size distribution of records $p(\Delta_r)$ is presented both for the original and 
 surrogate data sets. In all cases the bold lines represent best fits with 
 Eq.\ (\ref{eq:powdist}) obtained with the parameter values: 
 $\xi =2.15$, $c=1.0$ and $\Delta^*=1810$ for $p(\Delta)$,
 $\xi_r=1.0$, $c_r=0.9$ and $\Delta_r^*=1052$ for $p(\Delta_r)$, and
 $\xi_r^s=0.8$, $c_s=0.9$ and $\Delta_s^*=1270$ for  
 for shuffled series.
 }
 \label{fig:recsizedist}
 \end{figure}
Recently, we have shown that the size distribution of bursts $p(\Delta)$ 
accumulating 
all events up to failure during the compression process
has a power law functional form followed by an exponential
cutoff
\beq{
p(\Delta) \sim \Delta^{-\xi}\exp{\left[-(\Delta/\Delta^*)^c\right]}.
\label{eq:powdist}
}
In Refs.\ \cite{PhysRevE.88.062207,PhysRevLett.112.065501} 
the exponent $\xi$ was obtained numerically as $\xi=2.15$ for samples
comprising 20000 particles. In the present study $10^5$ particles are
used which gives the same exponent for $p(\Delta)$, however, with a broader
power law regime in Fig.\ \ref{fig:recsizedist}. 
Thus there is no evidence of scale-dependence in the power-lay exponent, 
only an increase in the characteristic size $\Delta^*$, and hence 
in the bandwidth of the power-law behaviour.

Records form a subset of events of the complete series containing solely
the largest bursts that occurred up to a given event index. 
Figure \ref{fig:recsizedist} shows that the distribution of the size of records 
$p(\Delta_r)$ has the same functional form Eq.\ (\ref{eq:powdist}) as the 
size distribution of all events $p(\Delta)$, 
however, the exponent $\xi_r$ of the power law regime is significantly 
lower $\xi_r=1.0\pm 0.05$ than for the complete distribution $\xi$. 
Selecting RB events implies a resampling of the ensemble of bursts, where 
the distribution of the record size is governed by the statistics of extremes. 
The difference in slope is likely caused by the inherent sample bias in choosing 
record breaking events either $(a)$ because this preferentially filters out smaller 
events and/or $(b)$ larger events may be associated with a greater 
degree of local stress concentration (stress intensity), known 
to be associated with a flatter slope \cite{sammonds_role_1992}.
In the shuffled event series large size events can occur earlier than
during the original fracture process so that small size events have a lower 
chance to become a record. As a consequence, the size distribution of shuffled
records has the same functional form, however, with a lower exponent 
$\xi_r^s=0.8\pm0.07$, which  indicates the lower fraction of small records.
 
\section{Approach to failure through breaking of records}
\begin{figure}%[!h]
 \begin{center}
 \epsfig{ bbllx=30,bblly=170,bburx=740,bbury=760,file=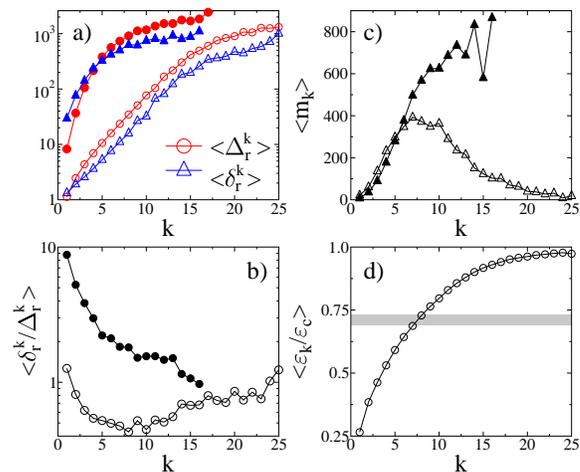,
 width=7.5cm}
 \end{center}
 \caption{%\small 
 (Color online) Average value of the characteristic quantities of records 
 as a function of the record 
 rank $k$: size and increment $(a)$, relative increment $(b)$, waiting time 
$(c)$,  and strain of records normalized by the critical strain of failure $(d)$.
 The Grey stripe highlights the strain range between the records of rank $k=7$ 
and $k=8$. In $(a), (b), (c)$ the open and filled symbols represent the original and 
the surrogate data sets, respectively.
 }
 \label{fig:averages}
 \end{figure}

The integrated statistics of records presented above gives an overall 
description of the subset of RB events of the crackling time series.
It is a question of fundamental interest how the system approaches failure
through a sequence of record breaking events.
To obtain a quantitative understanding we calculated averages 
of characteristic quantities of records as a function of the record rank $k$.
Figure \ref{fig:averages}$(a)$ shows that as the system evolves the average 
size of records $\left<\Delta_r^k \right>$ rapidly increases 
with $k$. The size increments $\left<\delta_r^k \right>$ 
exhibit qualitatively the same behavior, i.e.\ during the entire failure 
process
records get broken with an increasing sequence of increments. 
The shuffled data set has the same qualitative tendencies with the 
difference that low rank records reach higher sizes than in the original
data.
 
For a stationary sequence of IIDs the extreme value statistics leads 
to a monotonous decrease of average relative increments between consecutive records 
\cite{miller_scaling_2013}. 
This is what we find for the early part of our time series, when the timing 
and size of events is dominated by the random structural disorder.  
However, for the later RB events we find the opposite, i.e.\ an accelerating trend, 
most likely associated with processes dominated by the dynamics, such as stress 
relaxation and redistribution, and localization of events on to the eventual 
failure plane. This change from decelaration to acceleration is illustrated in 
Figure \ref{fig:averages}$(b)$ where the average
value of the relative increment $\left<\delta_r^k/\Delta_r^k \right>$ is 
presented. Initially the RB process slows down, i.e.\
up to a characteristic record rank $k^*=7-8$ the relative increment 
$\left<\delta_r^k/\Delta_r^k \right>$ decreases, while for 
higher ranks $k>k^*$ the onset of accelerating fracture is marked by the 
increase of $\left<\delta_r^k/\Delta_r^k \right>$.
The result is also supported by the behaviour of the surrogate data where 
correlation are destroyed: the relative increment monotonically decreases 
without any sign of qualitative change.
\begin{figure}%[!h]
 \begin{center}
 \epsfig{ bbllx=50,bblly=20,bburx=380,bbury=330,file=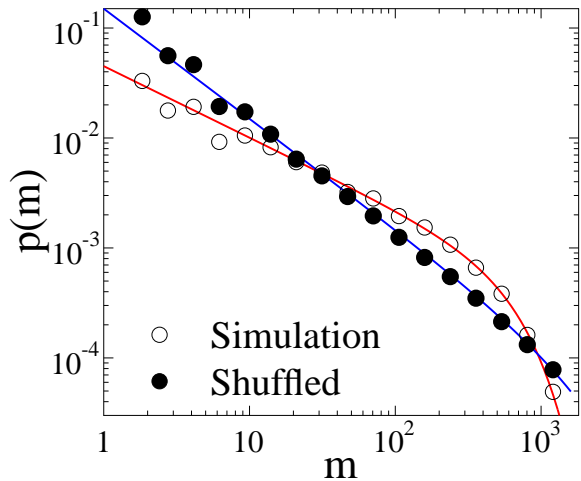,
 width=7.5cm}
 \end{center}
 \caption{%\small 
 Distribution $p(m)$ of waiting times $m$ between consecutive records for
 fracture simulations and for the surrogate data.
 The bold lines represent best fit which was obtained using the generic functional 
 form of Eq.\ (\ref{eq:powdist}).
 The value of the power law exponent is $z=0.62$ and $z=0.98$ for the simulated 
 and shuffled data, respectively.
 }
 \label{fig:waitdist}
 \end{figure}

After a record occurred as the $n_k$th event of the sequence
it gets broken by the next record which is the $n_{k+1}$th burst
of the evolving system. The waiting time $m_k=n_{k+1}-n_k$ 
is an important characteristics of the dynamics, it provides the number
of events one has to wait to break the $k$th record by the $(k+1)th$ one. 
The quantity $m_k$ 
is also called as the lifetime of the $k$th record.
For IIDs it has been
shown analytically that the probability distribution $p(m)$ of waiting times $m$
has a power law behavior 
\beq{
p(m) \sim m^{-z}
}
with a universal exponent $z_{IID}=1$. 
Figure \ref{fig:waitdist} demonstrates that 
the lifetime distribution of the surrogate is consistent with the IID
prediction with an exponent $z_s=0.98\pm 0.07$. For the fracture process 
the same functional form is obtained, however, 
with a different exponent $z=0.62\pm 0.03$. 
The low exponent $z<z_{IID}$ 
reflects the fact that during compressive failure of heterogeneous
materials long waiting times between records more frequently occur than for a 
random 
sequence of IIDs. Since record breaking is controlled by the statistics of 
extremes, 
in a sequence of IIDs waiting times get larger with the record rank $k$, since 
it takes 
longer to break a larger record. However, in the fracture process large waiting 
times 
occur at the beginning, in spite of their larger size, records of higher rank
may get broken faster because of the rapid increase of burst size when 
approaching failure.

The emergence of a characteristic record rank $k^*$ is further supported 
by the behavior of the average value of waiting times $m_k$. 
Figure \ref{fig:averages}$(c)$ presents the remarkable result that the record 
rank $k^*=7$ separates two qualitatively different regimes of the time evolution
of the compressed system: at the beginning of the failure process $k<k^*$ 
the increasing waiting time implies a deceleration of record breaking, where
it takes longer and longer to break the growing records. 
\begin{figure}%[!h]
 \begin{center}
 \epsfig{ 
bbllx=20,bblly=10,bburx=370,bbury=330,file=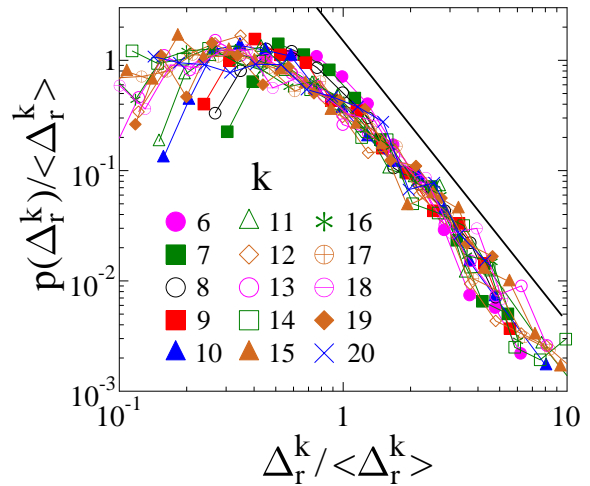,
 width=7.5cm}
 \end{center}
 \caption{%\small 
 (Color online) Probability distribution $p(\Delta_r^k)$ of record sizes 
$\Delta_r^k$  for fixed ranks $k$ rescaled with the average record size presented 
in Fig.\ \ref{fig:averages}$(a)$. Good quality collapse is obtained.
 }
 \label{fig:distreck}
 \end{figure}
For IIDs the average waiting time must be monotonically increasing 
\cite{PhysRevE.87.052811} as observed in Fig.\ \ref{fig:averages}$(c)$
for the surrogate data set 
since records can be overcome after a larger and larger number of trials drawn 
from the 
same distribution. Comparing our simulation results to the IID findings,
the increasing regime of $\left<m_k \right>$ can be attributed to the dominance 
of disorder 
in the fracture process consistent with our earlier findings 
\cite{PhysRevE.88.062207,PhysRevLett.112.065501}. 
Beyond $k^*$ the waiting time starts to decrease confirming the change of the 
dynamics. 
In spite of their larger size, records get broken after fewer 
and fewer small sized bursts which indicate the dominance of temporal and 
spatial 
correlations in triggering consecutive events. 
% \begin{figure}%[!h]
%  \begin{center}
%  \epsfig{bbllx=40,bblly=30,bburx=370,bbury=330,file=distrib_recsize_k_revise.eps,
%  width=7.5cm}
%  \end{center}
%  \caption{%\small 
%  (Color online) Probability distribution $p(\Delta_r^k)$ of record sizes 
% $\Delta_r^k$  for fixed ranks $k$. 
%  }
%  \label{fig:distreck_orig}
%  \end{figure}

In Ref.\ \cite{PhysRevE.88.062207} our simulations revealed the localization
of breaking bursts to a damage band, which occurs at a characteristic strain 
value close to failure. Such localization has also been seen in acoustic emission 
data during laboratory experiments loaded at constant strain rate \cite{sabine_ian_pre_2014}.
The spatial localization is accompanied by a rapid
increase of the average burst size 
\cite{PhysRevE.88.062207,PhysRevLett.112.065501}. 
In Fig.\ \ref{fig:averages}$(d)$ 
the average strain $\left<\varepsilon_k/\varepsilon_c\right>$ where the records 
occurred is presented normalized by the critical strain of macroscopic failure.
The strain range $\left<\varepsilon_k/\varepsilon_c\right> \approx 0.69-72$
corresponding to the record ranks $k^*=7-8$ is also highlighted
in the figure. Comparing to the strain of localization 
$\left<\varepsilon_k/\varepsilon_c\right> \approx 0.85-0.9$ it follows that 
the acceleration of the RB process sets in significantly earlier.  
The macroscopic response of the system proved to be quasi-brittle, i.e.\ 
linearly elastic behavior is obtained where stronger non-linearity emerges 
solely
close to failure due to the intensive bursting activity 
\cite{PhysRevE.88.062207,PhysRevLett.112.065501}. The yield stress 
$\sigma_Y$ and the corresponding strain $\varepsilon_Y$, which mark 
the onset of non-linearity of the constitutive curve $\sigma(\varepsilon)$, 
could be identified by computer simulations as 
$\sigma_Y/\sigma_c \approx 0.73$ and $\varepsilon_Y/\varepsilon_c \approx 0.7$,
respectively. The characteristic strain 
$\left<\varepsilon_{k^*}/\varepsilon_c\right>$
of accelerated record breaking falls  close to $\varepsilon_Y$.

To get a more detailed characterization of the evolution of the system 
towards failure, we evaluated the size 
distribution of records for fixed ranks $p(\Delta_r^k)$. 
Figure \ref{fig:distreck} presents a
 scaling plot where distributions of different record ranks are rescaled with 
the average
 record size $\left<\Delta_r^k\right>$ presented in Fig.\ 
\ref{fig:averages}$(a)$. 
Except for the first few bins of the smallest record sizes good quality data 
collapse is obtained which implies the validity of the scaling structure
 \beq{
 p(\Delta_r^k)=\left<\Delta_r^k \right>\Phi(\Delta_r^k/\left<\Delta_r^k 
\right>),
 }
where $\Phi(x)$ denotes the scaling function. The result demonstrates that 
records 
of different ranks have the same size distribution, they get only shifted to 
accommodate
the increasing average size. Note that the functional form of the scaling 
function  
$\Phi(x)$ can be approximated as a power law for $x>1$. 
The slope of the straight line in Fig.\
\ref{fig:distreck} is 2.55.
In the scaling analysis records of the lowest and highest
ranks are not included because their size spans only a narrow range, and
we do not have a sufficient number of events, respectively.
When the event series is shuffled, any burst can be the first event, and 
hence, the first record. However, for higher ranks small size events have 
a very low chance to become a record.
As a consequence the surrogate data set cannot have such a 
scaling behaviour because the functional form of the size distribution 
changes with the record rank: the size distribution of the first record 
is practically identical with the complete size distribution of bursts 
(see Fig.\ \ref{fig:recsizedist}), while for higher record ranks the 
distribution must have a slower decay.

\begin{figure}%[!h]
 \begin{center}
 \epsfig{ 
bbllx=10,bblly=10,bburx=380,bbury=330,file=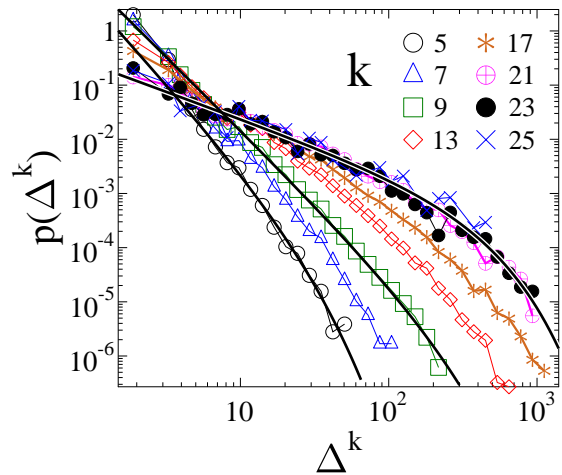,
 width=7.8cm}
 \end{center}
 \caption{%\small 
(Color online) Probability distribution $p(\Delta^k)$ of burst sizes in 
sub-sequences 
of events between consecutive 
records. For demonstration three curves were fitted with the functional form 
Eq.\ (\ref{eq:fit}). For the highest ranks $k\geq 20$ the distributions practically 
fall on the top of each other.
 }
 \label{fig:distsubseq}
 \end{figure}

\begin{figure}%[!h]
 \begin{center}
 \epsfig{ bbllx=10,bblly=10,bburx=350,bbury=300,file=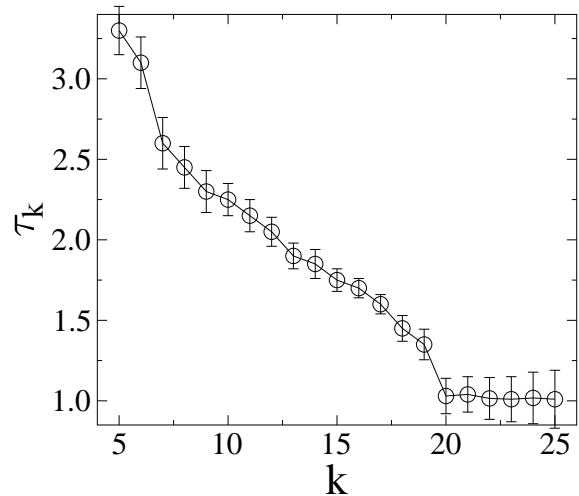,
 width=7.8cm}
 \end{center}
 \caption{(Color online) The value of the exponent $\tau_k$ of the size 
distribution of sub-sequences
 of bursts between consecutive records as a function of the record rank $k$,
 together with the error bars. 
 }
 \label{fig:expo}
 \end{figure}

In Fig.\ \ref{fig:series} it has been highlighted that consecutive records 
enclose subsequences of the complete time series. To explore this further 
we analyzed how the size distribution $p(\Delta^k)$
of these sub-sequences between the $k$th and $(k+1)$th records
evolve as the system approaches failure. We excluded the two record
bursts at the left and right hand sides of the sub-sequence from 
the statistics.
Figure \ref{fig:distsubseq} presents the resulting size distributions $p(\Delta^k)$ for 
several $k$ 
except for the lowest ranks ($k<5$) where event sizes span only a very narrow 
region 
so that no meaningful distributions could be obtained.
The distributions cannot be collapsed by rescaling with the average, only the 
cutoff of
the distributions could be scaled on the top of each other. The reason is that 
the distributions 
have a power law functional form followed by an exponential cutoff, 
\beq{
p(\Delta^k) \sim (\Delta^k)^{-\tau_k} e^{-\Delta^k/\Delta^k_s}
\label{eq:fit}
}
however, the power law exponents $\tau_k$ are different for different record 
ranks $k$. 
It can also be observed in Fig.\ \ref{fig:distsubseq} that the cutoff scale 
$\Delta^k_s$ grows with the record rank $k$ and the curves of the highest ranks
$k\geq 20$ fall practically on the top of each other. Figure \ref{fig:expo} presents
the exponent $\tau_k$ obtained by fitting with Eq.\ (\ref{eq:fit}). 
For the lowest rank the exponent has a high value $\tau_k = 3.26$ then 
it monotonically decreases and for the highest ranks it tends to the vicinity of
$\tau_k = 1.0$.

The result is an interesting manifestation of the b-value anomaly 
\cite{bvalue_Scholz01021968}
(i.e. the change in the exponent of the log-linear frequency-magnitude distribution)
we have pointed out
before in DEM simulations of compressive failure \cite{PhysRevE.88.062207}.
In sub-sequences between records 
the number of events is practically equal to the waiting time $m_k$, 
which covers a broad range (see Fig.\ \ref{fig:averages}$(b)$). 
In the traditional analysis of the time series of bursts, event windows are 
considered 
with a fixed number (typically a few hundred) of events. This has the consequence
that windows close to failure involve more and more records and sub-sequences
which results in an exponent different from the one of single sub-sequences 
presented above
\cite{PhysRevE.88.062207,PhysRevLett.112.065501}.
 \begin{figure}%[!h]
 \begin{center}
 \epsfig{ 
bbllx=40,bblly=30,bburx=370,bbury=330,file=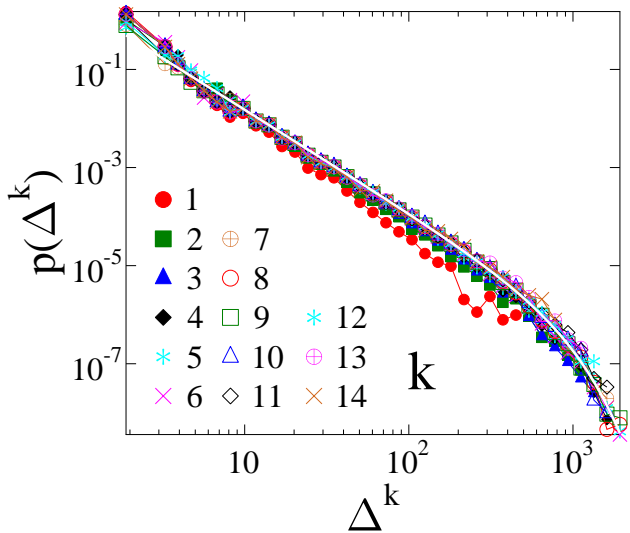,
 width=7.8cm}
 \end{center}
 \caption{%\small 
(Color online) Probability distribution $p(\Delta^k)$ of burst sizes in 
sub-sequences of the shuffled event series between consecutive 
records. Except for the lowest ranks the curves fall on the top of each other.
The white curve was obtained by fitting the functional form Eq.\ (\ref{eq:fit})
with an exponent $\tau=2.1$ nearly equal to the one of the size distribution including
all the bursts.
}
 \label{fig:distsubseq_shuff}
 \end{figure}
 
Figure \ref{fig:distsubseq_shuff} shows that no such behaviour exists when correlations
are destroyed by shuffling the event series. Except for the lowest ranks, practically 
all the distributions of the sub-sequences fall on the top of each other. The emerging 
master curve can be well fitted with the functional form Eq.\ (\ref{eq:fit}) 
where the exponent takes the value $\tau=2.1$ falling close to the size distribution exponent
of the whole population of the bursts. The result is reasonable since all the sub-sequences
are random samples of the complete event set.

\section{Structure of the time series}
\begin{figure}%[!h]
 \begin{center}
 \epsfig{ bbllx=20,bblly=30,bburx=390,bbury=330,file=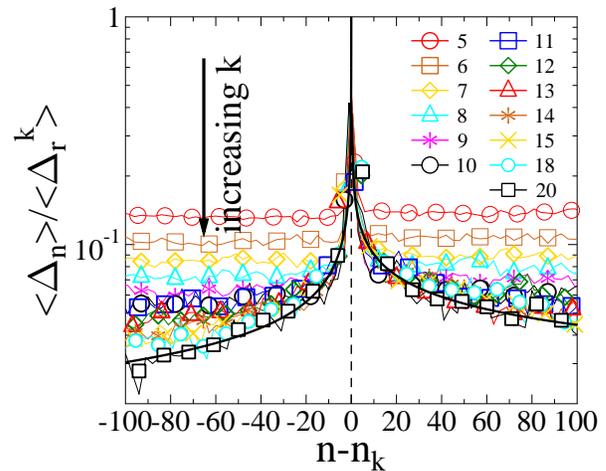,
 width=8.0cm}
 \end{center}
 \caption{%\small 
(Color online) Average size of events $\left<\Delta_n\right>$
in the vicinity of records as a function of their integer index relative 
to records $n-n_k$.
High rank records are preceded by events of increasing size and they are 
followed by a relaxation process where the burst size gradually decreases.
The bold line represents fit with Eq.\ (\ref{eq:averburst_index}).
}
 \label{fig:apprecburstscaled}
 \end{figure}

Records are major events of the breaking activity of the compressed sample, 
which can 
have a strong effect on both the spatial and temporal occurrence of 
subsequent events.
In Figure \ref{fig:series} the increasing value of the moving average of the 
burst size 
towards records may indicate some kind of precursory activity preceding record 
breaking 
events. In order to obtain a more detailed description of the internal structure 
of the 
time series we calculated the average size of bursts $\left<\Delta_n\right>$ 
before and after records of different 
ranks $k$ as a function of the event index $n-n_k$ relative to the records 
$n_k$. 
In Figure \ref{fig:apprecburstscaled} zero index corresponds to the records 
while 
positive and negative
values of $n-n_k$ stand for events preceding and following records, 
respectively.
For clarity, the $\left<\Delta_n\right>$ curves are normalized by the average 
size 
$\left<\Delta_r^k\right>$ of the corresponding record.
 
It can be observed in Fig.\ \ref{fig:apprecburstscaled} that low rank records 
just randomly pop up on a flat background without any signature of the imminent 
record
event. However, high rank records are approached through a sequence of 
precursory events with an
increasing average burst size, and they are followed by a relaxation process 
characterized by a
gradually decreasing burst size. 
In this sense the later record-breaking events mimic the behaviour 
of the final dynamic rupture event.
The functional form of $\left<\Delta_n\right>$ 
can be well fitted by an inverse power law 
\beq{
\left<\Delta_n\right> \sim \left| n-n_k \right|^{-\gamma} 
\label{eq:averburst_index}
}
on both sides of high rank records. In Figure \ref{fig:apprecburstscaled} 
a best fit is obtained with the exponents 
$\gamma=0.38$ and $\gamma=0.31$ before and after records, respectively, 
considering
the curves of $k=18,20$.
 
\begin{figure}%[!h]
 \begin{center}
 \epsfig{ 
bbllx=30,bblly=30,bburx=380,bbury=560,file=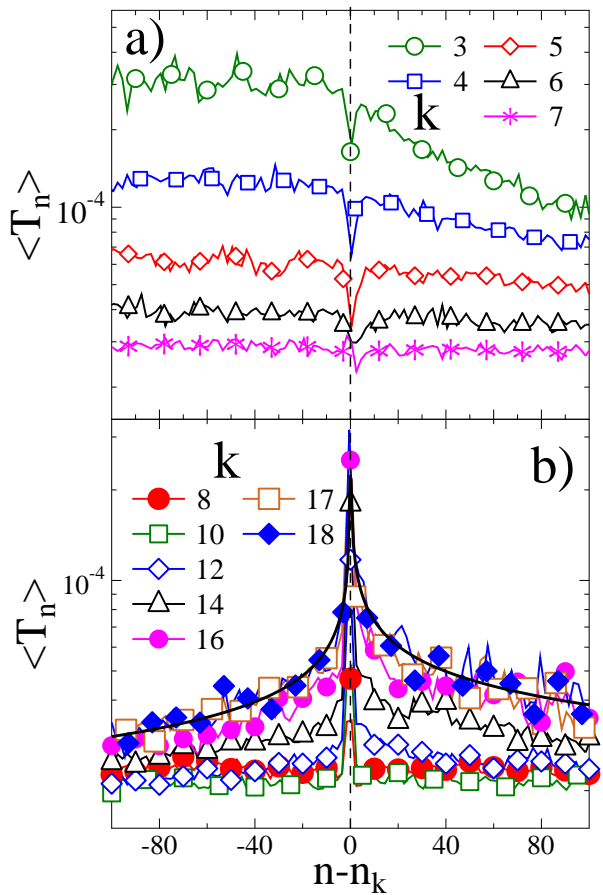,
 width=8.0cm}
 \end{center}
 \caption{%\small 
(Color online) Average value of the physical waiting time $\left<T_n\right>$ 
between consecutive 
events as a function of the event index relative to records $n-n_k$. The low 
and 
high rank records are presented separately in $(a)$ and $(b)$, respectively. 
The 
red bold continuous lines in $(b)$ represent fit with Eq.\ 
(\ref{eq:averburst_index}).
}
 \label{fig:apprecwait}
 \end{figure}

To characterize the temporal occurrence of events we analyzed in a similar way 
the 
average value of the physical waiting time $\left<T_n\right>$ between 
consecutive events 
before and after records as a function of the event index $n-n_k$ relative to 
record 
indices $n_k$. Figure \ref{fig:apprecwait} shows that for low record ranks $k<7$ 
the 
$\left<T_n\right>$ curves have a sharp local minimum at records which implies 
that the records 
are approached by a short period of increasing event rate and they are followed 
by a few relaxation events with increasing waiting times in between. However,
for higher record ranks $k\geq 7$ the overall behavior drastically changes: 
The $\left<T_n\right>$ curves develop a broader and broader maximum at records 
which shows that large size
events occurring in the vicinity of records (compare also to Fig.\ 
\ref{fig:apprecburstscaled}) 
are followed by longer waiting times. This behavior is the consequence of the 
strain controlled
loading, i.e.\ after large size events it takes longer to build up again the 
stress field and initiate the next bursts. In a different context, 
studying the correlation of event size and the waiting time after events 
we have found analogous behavior in Ref.\ 
\cite{PhysRevE.88.062207,PhysRevLett.112.065501}.
The present result is another form of appearance of 
the size-waiting time correlation. 
In the slow-down and acceleration periods the $\left<T_n\right>$ curves can be 
well
fitted with the power law functional form. For the exponents providing best fit 
in Fig.\ \ref{fig:apprecwait}
the same value 0.29 was obtained numerically on both sides of records. 

\section{Discussion}
We investigated the statistics of record breaking bursts generated during the 
compressive 
failure of heterogeneous materials. A synthetic time series of crackling events 
was 
generated by discrete element simulations of a realistic model where a 
cylindrical
sample was subject to uniaxial compression in a strain controlled way. 
Records of the time series of crackling events are identified as breaking 
bursts 
with size greater than any previous burst. The large system size 
in the model and the 
large number of samples generated by simulations allowed us to obtain high 
quality results for the statistics of records. In order to reveal the role of correlations
of bursts the analysis was repeated for a surrogate time series which was 
generated by randomly reshuffling the simulated events. Additionally, our results were 
compared to the corresponding analytic findings on sequences of independent 
identically distributed random variables, as well.

The overall statistics of records is characterized by power law distributions
of the size and lifetime of records and by the power law increase of the record 
number with the total number of events. This behavior deviates from 
the surrogate data set where correlations were destroyed, however, the 
later one agreed very well with the stationary process of IIDs. 

As the compression proceeds the system gradually evolves towards macroscopic 
failure where the sample loses its integrity. We have shown earlier 
\cite{PhysRevLett.112.065501,PhysRevE.88.062207} that the failure 
process has two qualitatively different stages: the beginning of the 
failure process is dominated by the structural disorder of the sample giving 
rise to an uncorrelated emergence of small sized breaking bursts. Later on as 
macroscopic failure is approached the fracture process accelerates which 
is indicated by the 
increasing burst size and by the decreasing waiting times between consecutive 
bursts. This second stage is dominated by the growing spatial and temporal 
correlations of local breaking events in the sample. 
In the present paper we showed that analyzing
the statistics of record breaking bursts and their evolution with increasing 
rank, provides an alternative way for the quantitative characterization 
of the emerging correlations in the fracture process as the loaded system evolves 
towards failure. 

Our analysis revealed the existence of a characteristic record rank $k^*$ which 
separates
two regimes of the fracture process: for low record rank $k<k^*$ the process of 
record breaking slows down, i.e.\ it takes longer and 
longer to break a record. At $k^*$ the RB process starts to accelerate indicated by 
the rapidly decreasing record lifetime and by the increasing relative size 
increments of records. The surrogate event series proved to follow the IID predictions
without any sign of the emergence of a characteristic record rank.

Records are found to affect the surrounding structure of the 
time series: approaching high rank records, bursts have a gradually increasing 
size separated by an increasing physical waiting time, while after records the event 
size decreases and the temporal occurrence accelerates with decreasing 
waiting times. The evolution of both the event size and waiting time is characterized 
by power law functional forms as a function of the distance from records.
The result is consistent with our earlier finding presented in 
Ref.\ \cite{PhysRevLett.112.065501}, where the increase of waiting time after 
large size events has been pointed out by analyzing the 
size-waiting time correlation of bursts. The reason is
the strain controlled loading, which has the consequence that larger events release 
the stress in a larger volume of the specimen, and hence, it takes longer to 
build up the stress field again and trigger the next burst. 

Scaling analysis revealed that records of fixed ranks have the same size 
distribution which has power law asymptotics. Records split the time 
series of bursts into sub-sequences which typically contain fewer 
and fewer events as the system approaches
macroscopic failure. We showed that bursts of sub-sequences are characterized 
by power law size distribution, however, the exponent spans a broad range 
decreasing to the vicinity of one towards failure. 
This result is consistent with the trend of 
the b-value anomaly of the time series \cite{sammonds_role_1992}.
In traditional b-value analysis of crackling time series windows of events
are considered either with a fixed number of events or with a fixed strain or 
stress increment. Since records affect the structure of the time 
series, our analysis suggests that focusing on subsequences between records 
can provide significant additional and relevant information to that which 
can be inferred from the average properties of the time series of the whole 
population of all events.  In principle this could improve forecasting power 
for catastrophic failure, but this depends on our being able to detect the 
processes revealed here in real time, and in a single realization, 
during a 'live' experiment. 

Our simulations were carried out with samples of about 100000 particles 
which corresponds to practically laboratory scale sample sizes. Comparison
to simulations of a significantly smaller system size of 20000 particles 
(see Refs.\ \cite{PhysRevLett.112.065501,PhysRevE.88.062207}) 
showed that in larger systems a larger number of crackling bursts are generated 
whose size spans a broader range. However, the probability distributions 
of the characteristic quantities of the event ensemble are quite robust, 
i.e.\ the value of the power law 
exponents are the same within the error bars. The same is valid for the number 
and size of records, however, for the value of the characteristic record rank 
$k^*$ no size dependence could be pointed out. 
In the simulations with 100000 particles the relative fluctuation of the total 
number of bursts is quite moderate (below 0.1). Tests of the effect of these
fluctuations, e.g.\ by restricting the analysis to simulations where 
the number of bursts exceeds a threshold, revealed that they mainly affects the 
statistics of the results but no systematic bias occurred.

Our analysis strongly relies on the comparison of the record statistics of 
crackling noise time series to the corresponding IID findings and to the results 
of uncorrelated surrogates.
A similar strategy has recently been applied in Ref.\ \cite{PhysRevE.87.052811}
to study the avalanche dynamics of models of self organized criticality where IID 
results were derived by substituting the known steady state distribution of 
avalanche sizes into the generic IID formulas. Deviations of the record statistics of 
SOC models from the IID results were identified as signatures of 
correlations emerging in the dynamics of avalanches. 
Record breaking statistics of daily temperatures has proven a useful tool 
to investigate the effect of global warming. 
It has been pointed out in Ref.\ \cite{PhysRevE.74.061114} that the 
observed frequency and average value of record temperatures 
can be understood in terms of a stationary climate, so that the current warming rate does not 
have a noticable effect on the record statistics of daily temperatures (at least 
in the city of Philadelphia where the data was measured). 
Additionally, the ratio of the number of record high 
(record of the highest temperature of a given day) and record low 
(record of the lowest temperature of a given day) temperatures 
was suggested as a useful measure to point out trends in time series 
\cite{PhysRevE.74.061114}.
This idea was extended to earthquakes by Ref.\ \cite{npg-17-169-2010} where the 
sequence of interval times between successive earthquakes were analyzed. It was 
shown that for global earthquakes the statistics of both record breaking longer and 
record breaking shorter intervals is consistent with IID processes. 
However, for an isolated aftershock sequence, where 
inter event times get systematically longer, the number of record breaking longer
intervals was found to increase as a power law of the event number similarly to our result 
Eq.\ (\ref{eq:averrwcnum}).  
At the same time the ratio $r$ of the number of record breaking 
longer and shorter intervals proved to be predominantly greater than one, which
was suggested as a measure to distinguish between background seismicity 
and aftershock sequences \cite{npg-17-169-2010}. Extending the calculation of the ratio $r$
to a broader spatiotemporal interval, before and after the mainshock, $r<1$ and $r>1$ 
were found, respectively, which addressed a possibility of using record breaking 
statistics for forecastig \cite{npg-17-169-2010}.
For inter event times of aftershocks the Omori-type 
relaxation was proved to be responsible for the power law increase of the number 
of records, while in our case
the monotonically increasing average burst size towards failure plays a similar role.
For earthquakes, the frequency-magnitude distributions
for main shocks and aftershocks are the same, however, for fracture processes 
both the event magnitude and inter-event time exhibit a systhematic evolution
towards macroscopic failure. In the present study we focused on the record breaking 
process of event magnitudes, the extension to inter-event times is a work in progress
including also the analysis of the correlation of consecutive records
\cite{krug_PhysRevLett.108.064101}.

\begin{acknowledgments}
The work is supported by the projects TAMOP-4.2.2.A-11/1/KONV-2012-0036.
The project is implemented through the New Hungary Development Plan,
co-financed by the European Union, the European Social Fund and 
the European Regional Development Fund.
F.\ Kun acknowledges the support of OTKA K84157.
\end{acknowledgments}

\bibliography{/home/feri/papers/statphys_fracture}

\end{document}